# MESSENGER observations of Mercury's planetary ion escape rates and their dependence on true anomaly angle

Weijie Sun[1]* (weijiesun@berkeley.edu), Ryan M. Dewey[2], Xianzhe Jia[2], Jim M. Raines[2], James A. Slavin[2], Yuxi Chen[3], Tai Phan[1], Gangkai Poh[4,5], Shaosui Xu[1], Anna Milillo[6], Robert Lillis[1], Yoshifumi Saito[7], Stefano Livi[2,8], Stefano Orsini[6]

[1]Space Sciences Laboratory, University of California, Berkeley, CA 94720, USA

[2]Department of Climate and Space Sciences and Engineering, University of Michigan, Ann Arbor, MI 48109, USA

[3]Center for Space Physics and Department of Astronomy, Boston University, Boston, MA 02215, USA

[4]NASA Goddard Space Flight Center, Greenbelt, MD, USA

[5]Center for Research and Exploration in Space Sciences and Technology II, Catholic University of America, Washington, DC, USA

[6]Institute of Space Astrophysics and Planetology, INAF, Rome, Italy

[7]Japan Aerospace Exploration Agency, Institute of Space and Astronautical Science, Sagamihara, Japan

[8]Southwest Research Institute, San Antonio, TX, USA

**Running Title:** $Na^+$-group Ions Escape at Mercury

**Key Points**:
1, $Na^+$-group ions form escape plumes in solar wind and magnetosheath, with higher fluxes in the positive solar wind electric field hemisphere
2, The escape rate ranges from 0.2 to $1 \times 10^{25}$ atoms/s, with the magnetosheath being the main escaping region
3, Escape rates peak near perihelion, and are similar during other true anomaly angle intervals

**Abstract** (<150 words).
This study investigates the escape of Mercury's sodium-group ions ($Na^+$-group, including ions with m/q from 21 to 30 amu/e) and their dependence on true anomaly angle (TAA), i.e., Mercury's orbital phase around the Sun, using measurements from MESSENGER. The measurements are categorized into solar wind, magnetosheath, and magnetosphere, and further divided into four TAA intervals. $Na^+$-group ions form escape plumes in the solar wind and magnetosheath, with higher fluxes along the solar wind's motional electric field. The total escape rates vary from 0.2 to $1 \times 10^{25}$ atoms/s with the magnetosheath being the main escaping region. These rates exhibit a TAA dependence, peaking near the perihelion and similar during Mercury's remaining orbit. Despite Mercury's tenuous exosphere, $Na^+$-group ions escape rate is comparable to other inner planets. This can be attributed to several processes, including that $Na^+$-group ions may include several ion species, efficient photoionization frequency for elements within Na-group, etc.

**Plain Language Summary** (<200 words)
Atmospheric escape is defined as the loss of atmospheric particles in the form of neutrals and ions into outer space. Understanding atmospheric escape is a fundamental science question for studying the evolution of planetary atmosphere and habitability, as it can provide insight into how life can form on a planet. While atmospheric escape has been extensively studied in Venus, Earth, and Mars through *in situ* measurements and simulations, it remains poorly understood at Mercury. In this study, we investigate the escape of the most abundant heavy ions at Mercury, specifically the $Na^+$-group ions, using MESSENGER's measurements. Our findings show that the escape rates of the $Na^+$-group ions are dependent on Mercury's orbital phase around the Sun, exhibiting a seasonal effect with rates from 0.2 to $1 \times 10^{25}$ atoms/s. This rate is comparable to the ion's escape rates at other inner planets, which is surprising given that Mercury only has a tenuous exosphere. We propose that this can be attributed to several processes such as efficient photoionization, solar wind sputtering, and solar wind momentum exchange at Mercury, and the $Na^+$-group ions include several ion species such as $Na^+$, aluminum ion ($Al^+$), silicon ion ($Si^+$) and magnesium ion ($Mg^+$) etc.

1. Introduction

Understanding how atmospheric escape from a planet is crucial for studying atmospheric evolution and planetary habitability. Atmospheric escape can occur through thermal escape (thermally-driven escape and atmospheric heating-driven hydrodynamic escape) and non-thermal escape (photochemical-induced neutral escape and plasma-related ion escape) (e.g., Gronoff et al., 2020; Johnson et al., 2008). Among these processes, ion escape is an important channel and has been extensively studied at Venus, Earth, and Mars (e.g., Ramstad & Barabash, 2021). For instance, the escape rate of $O^+$ ions at Mars (0.3 to $2 \times 10^{25}$ $s^{-1}$) (Dong et al., 2017; Lundin et al., 2009, 2013; Nilsson et al., 2011) can contribute up to one third of the total oxygen loss ($6 \times 10^{25}$ $s^{-1}$) (Jakosky et al., 2018).

Mercury is the planet closest to the Sun and has a thin exosphere (e.g., Broadfoot et al., 1974). Mercury's magnetosphere, formed through the interaction of its global intrinsic magnetic field with the solar wind, possesses a magnetopause approximately 1000 kilometers above the planet's surface near the subsolar point (e.g., Siscoe et al., 1975; Slavin et al., 2008). Mercury's exosphere contains a variety of neutral atoms such as hydrogen (H), oxygen (O), sodium (Na) and magnesium (Mg) (e.g., Domingue et al., 2007; Merkel et al., 2017; McClintock et al., 2018) and their abundance varies with Mercury's orbital phase around the Sun, i.e., seasonal variations (e.g., Cassidy et al., 2015, 2016; Lammer et al., 2003). Mercury's orbital phase is often denoted by its True Anomaly Angles (TAAs), representing the angle between the directions of Mercury's perihelion and its current position. Notably, Mercury's orbit is the most eccentric among planetary orbits in the solar system (see, Sun, Dewey, et al., 2022), ranging 0.31 to 0.47 astronomical unit (AU).

Previous studies have examined the escape rate of Na atoms, the most abundant heavy atom in Mercury's exosphere. These studies reveal that solar radiation can remove between 0.5 and $1.3 \times 10^{24}$ neutral Na atoms/s (McClintock et al., 2008; Potter & Killen, 2008; Schmidt et al., 2010). Photoionization and convection electric field pickup contribute to $Na^+$ ion loss with a range from $10^{23}$ atoms/s (Hunten et al., 1988; Leblanc & Johnson, 2003) to $10^{24}$ atoms/s (Ip, 1990) in models. Jasinski et al. (2021) calculate a range of 0.9 to $4 \times 10^{24}$ atoms/s with the maximum occurring at perihelion.

Measurements from Fast Imaging Particle Spectrometer (FIPS) (Andrews et al., 2007) onboard MESSENGER (Solomon et al., 2007) reveal that sodium group ions ($Na^+$-group ions) are the most abundant heavy ions near Mercury, with enhancements observed in the northern cusp, plasma sheet, and dawn terminator (Raines et al., 2013; Zurbuchen et al., 2011). Sun, Slavin, et al. (2022) identify that solar wind sputtering over the northern cusp can cause an enhance of $Na^+$-group ions at a rate of $10^{24}$ $s^{-1}$. However, a comprehensive *in situ* investigation of the $Na^+$-group ion escape rate remains to be conducted.

In this paper, we deliberately distinguish between $Na^+$ ion, which specifically refer to sodium ion, and $Na^+$-group ions, including ions with m/q from 21 to 30 amu/e such as $Na^+$ ion, aluminum ion (Al$^+$), silicon ion (Si$^+$) and magnesium ion (Mg$^+$). The modeling results described in this introduction are specifically about Na$^+$ ion. Our study focuses on Na$^+$-group ions. We analyze their distributions in the solar wind, magnetosheath, and magnetosphere under different TAA intervals, using MESSENGER's *in situ* measurements. We identify escape plumes in the solar wind motional electric field coordinate, and we estimate the escape rates and their seasonal variations across different regions.

## 2. Data Source and Instrumentation

This study utilizes magnetic field and ion measurements from MESSENGER, a 3-axis stabilized single spacecraft mission that orbited Mercury from 18 March 2011 to 30 April 2015, UTC. The magnetometer (Anderson et al., 2007) provided magnetic field measurements with a time resolution of up to 20 Hz in the Mercury solar orbital (MSO) coordinate system. Ions were measured by FIPS, a part of Energetic Particle and Plasma Spectrometer (EPPS).

FIPS was an ion-mass spectrometer that could resolve m/q from 1 to 60 atomic mass units (amu/e) using energy per charge (E/q) and time-of-flight measurements. The E/q spanned approximately 0.046 to 13.3 keV/e. FIPS achieved a time resolution of around 10 seconds below the bow shock and 67 seconds outside, with an effective field-of-view of approximately $1.15\pi$ sr and an angular resolution of 15°. The Na$^+$-group ions including ions with m/q from 21 to 30 amu/e to enhance signal-to-noise ratios (Raines et al., 2011). This study excludes ions with energy below 2 keV, which are under further investigation.

## 3. Observation

### 3.1. Distribution in Solar Wind and Magnetosheath

Figure 1 shows the distributions of Na$^+$-group ions upstream of Mercury's bow shock and within the mangetosheath under solar wind convection electric field coordinates (MSE). The measurement locations are initially converted from the MSO coordinate to the Mercury solar magnetospheric (MSM) coordinate, accounting for a northward offset of approximately 0.2 R$_M$ (Mercury radius) (Alexeev et al., 2008; Anderson et al., 2012), and then transformed into the MSE coordinates. In MSE coordinates, $\hat{x}_{MSE}$ is antiparallel to the solar wind velocity $(\vec{V}_{sw})$, $\hat{z}_{MSE}$ aligns with the motional electric field $(\vec{E} = -\vec{V}_{sw} \times \vec{B}_{IMF})$, and $\hat{y}_{MSE}$ completes the right-handed coordinate system. For measurements in the solar wind, $\vec{V}_{sw}$ is assumed to be along the $-\hat{x}_{MSM}$ direction, and $\vec{B}_{IMF}$ is obtained by averaging *in situ* magnetic field measurements within each FIPS scan. For the magnetosheath, MSE coordinates are established from $\vec{B}_{IMF}$ averaged 10 minutes upstream of the nearest bow shock crossing, and the $\vec{V}_{sw}$. This assumption is supported by statistical studies showing that the IMF's direction mostly remains similar downstream of the bow shock (James et al., 2017). MESSENGER's measurements are mostly taken from the southern hemisphere, the use of MSE coordinates may help mitigate this non-uniform spatial coverage (see supplementary material).

The particle fluxes of $Na^+$-group ions in the magnetosheath (Figure 1e) are several times of those in the solar wind (Figure 1b). Figures 1b and 1e show higher particle fluxes in the positive $E$ hemisphere than in the negative $E$ hemisphere. In the 3-dimensional (3D) distribution (Figures 1c and 1f), which accumulates measurements with $E_\theta$ from 40° to 60°, $Na^+$-group ions concentrate in the hemisphere of positive $E$, i.e., $+\hat{z}_{MSE}$, indicating that they are moving along the motional electric field. This feature of $Na^+$-group ions resembles the escaping plume ions observed at Mars (Carlsson et al., 2006; Dubinin et al., 2006; Dong et al., 2015), Venus (Dubinin et al., 2013; Xu et al., 2023) and the Moon (Halekas et al., 2012; Poppe et al., 2022).

### 3.2. Escape Rates through Solar Wind and Magnetosheath

To estimate escape rates of $Na^+$-group ions in the solar wind and magnetosheath, we integrated measurements from -3 to 0 $R_M$ in $X_{MSE}$ across four TAA intervals: 45° to 135°, 135° to 225°, 225° to 315°, and 315° to 45°. The selection of the integrated range considers ion injection into the magnetosphere, as elaborated in section 4.3. Figures 2a and 2d show the observed density of $Na^+$-group ions in the solar wind and magnetosheath, respectively, in the $\hat{y}_{MSE} - \hat{z}_{MSE}$ plane for the TAA interval from 225° to 315°. These figures reveal two torus-shaped regions, but are similar to hollow cylinders in 3D.

Figures 2b and 2e show the accumulated 3D distributions of $Na^+$-group ions for the measurements within a sector of Figures 2a and 2d, respectively. These distributions reveal a significant portion of particles moving tailward ($-\hat{x}_{MSE}$), which are escaping particles. The transport rates of $Na^+$-group ions are calculated by multiplying the particle fluxes by the covering areas. For the solar wind, the covering area is approximately 61 $R_M^2$, with an inner circle of radial distance of 3.1 $R_M$ and an outer circle of 5.5 $R_M$ for the torus region in Figure 2a. The sunward and tailward particle fluxes along $+\hat{x}_{MSE}$ and $-\hat{x}_{MSE}$, are $8.3 \times 10^3$ and $5.2 \times 10^4$ cm$^{-2}$ s$^{-1}$, resulting in transport rates of $1.2 \times 10^{23}$ and $7.6 \times 10^{23}$ s$^{-1}$ respectively, and a net tailward rate, or the escape rate, of $6.4 \times 10^{23}$ s$^{-1}$. Similarly, the sunward and tailward rates through the magnetosheath are calculated to be $8.1 \times 10^{22}$ and $1.3 \times 10^{24}$ s$^{-1}$, yielding an escape rate of $1.2 \times 10^{24}$ s$^{-1}$.

Figures 2c and 2f summarize the transport rates of $Na^+$-group ions along with the TAA intervals. It is evident that the majority of particles are moving tailward, and that the escape rates peak near the perihelion, i.e., TAA from 315° to 45°.

### 3.3. Escape Rates through Magnetotail

Figures 3a and 3b show the distributions of $Na^+$-group ions in the magnetotail in the $\hat{y}_{MSM} - \hat{z}_{MSM}$ plane for two TAA intervals: 315° to 135°, and 135° to 315°. Since ions dynamics inside the magnetosphere are not directly controlled by the solar wind $\vec{E}$, their distributions are shown under the MSM coordinate and they were divided into two TAA intervals for comprehensive magnetotail coverage.

Figure 3c presents the accumulated 3D distribution of Na$^+$-group ions under the MSM coordinates for the measurements of Figure 3a. This distribution reveals that a large portion of the particles moves tailward ($-\hat{x}_{MSM}$). The sunward and tailward particle fluxes are $3.9 \times 10^5$ and $1.4 \times 10^6$ cm$^{-2}$ s$^{-1}$. The covering area is approximately 28 R$_M^2$, resulting in sunward and tailward transport rates of $6.6 \times 10^{23}$ and $2.4 \times 10^{24}$ s$^{-1}$, with a net tailward rate of $1.8 \times 10^{24}$ s$^{-1}$. For the TAA interval from 135° to 315°, the sunward and tailward rates are $3.5 \times 10^{23}$ and $4.9 \times 10^{23}$ s$^{-1}$ with a net tailward rate of $1.4 \times 10^{23}$ s$^{-1}$ (see section 4.3 for a discussion on the ion escape within the magnetosphere).

### 3.4. Total Escape Rates and Variations with True Anomaly Angles

Figure 4a summarizes the escape rates of Na$^+$-group ions through the solar wind, magnetosheath and magnetotail. Na$^+$-group ions primarily escape through the magnetosheath, with escape rates several times those through the solar wind and the magnetotail during most of the TAA intervals. The maximum escape rate of Na$^+$-group ions ($10^{25}$ s$^{-1}$) occurs within the TAA interval from 315° to 45° including the perihelion. The escape rates during other TAA intervals, including outbound orbit (45° to 135°), near aphelion (135° to 225°), and outbound orbit (45° to 135°), are comparable, i.e., 2 to $3 \times 10^{24}$ s$^{-1}$.

## 4. Discussion

### 4.1. Energy of Pickup Ions and Energy Range of FIPS

Our study demonstrates that the Na$^+$-group ions, forming escape plumes and concentrating in the positive $\vec{E}$ direction in the solar wind and magnetosheath, are pickup ions. The solar wind pickup ions in the distribution function follow a circle centered on the solar wind ions with a radius equivalent to solar wind's perpendicular velocity component ($v_{\text{sw\_perp}}$). Thus, the velocity of these pickup ions ranges from zero and twice the $v_{\text{sw\_perp}}$ (see, Möbius et al., 1985; Gloeckler et al., 1993).

Given that the separation angle between the solar wind and IMF near Mercury's orbit is approximately 25°, along with a solar wind speed of around 350 km/s (Sun, Dewey, et al., 2022), the $v_{\text{sw\_perp}}$ is calculated to be 148 km/s. Therefore, the maximum velocity of pickup Na$^+$-group ions is anticipated to be approximately 296 km/s, i.e., 19 and 27 keV for atomic masses of 21 and 30, respectively. The energy coverage for the Na$^+$-group ions from FIPS was set between 2 and 13.3 keV, accounting for 43% to 61% of the total pickup energy range. This suggests that approximately half of the pickup ions were not included in our calculations.

### 4.2. TAA Dependence

Our results show that the escape rates of Na$^+$-group ions depend on the TAA, which corresponds to Mercury's orbital phase around the Sun. The escape rate peaks at $10^{25}$ s$^{-1}$ within the TAA interval from 315° to 45°, which includes the perihelion. This peak rate is much higher than the escape rates observed in other Mercury's orbital intervals: the outbound interval (TAA from 45° to 135°), near aphelion (135° to 225°), and the inbound interval (225° to 315°), all of which approximate $2 \times 10^{24}$ s$^{-1}$.

Despite the Na$^+$-group ions include multiple ion species, their maximum escape rate near perihelion is consistent with the calculated photoionization loss and the photoionization frequency of neutral Na (see Figure 3e in Jasinski et al., 2021). Jasinski et al. (2021) report a higher production rate of Na$^+$ ion during Mercury's inbound orbit compared to its outbound orbit. This production rate depends on the abundance of neutral Na in the exosphere, regulated by solar radiation intensity (Ip, 1986; Smyth & Marconi, 1995). However, our results show that the escape rates of Na$^+$-group ions are comparable during the inbound and outbound orbits. The discrepancy between our observations and the previous theoretical calculations could be partially attributed to the presence of multiple species in the Na$^+$-group ions. Furthermore, the escape of ions is also influenced by other processes including magnetospheric dynamics (see the following section).

### 4.3. Magnetospheric Dynamic Influence

Various physical processes in Mercury's magnetosphere could influence the dynamics of planetary ions. Firstly, magnetospheric dynamic may generate planetary ions through sputtering. Observations and simulations have revealed that magnetic reconnection on Mercury's dayside magnetopause enhances solar wind sputtering underneath the cusp (Poh et al., 2016; Raines et al., 2022; Fatemi et al., 2020), which cause enhancement of Na$^+$-group ions at high latitudes (Sun, Slavin, et al., 2022). This sputtering effect is expected to also occur on the surface beneath the nightside plasma sheet (Glass et al., 2022).

Secondly, planetary ions could recycle back to the planetary surface in the magnetosphere. Simulations of Na$^+$ ion indicate that in Mercury's magnetosphere they can return to the planetary surface at percentages ranging from < 15% to 70% (e.g., Ip, 1990; Killen et al., 2004; Leblanc & Johnson, 2010). However, ions escaping through the magnetosheath and solar wind are mostly expected to be lost downtail. Figure 4b shows the trajectories of several test Na$^+$ ions, which traced the global electromagnetic fields extracted from the Hall-Magnetohydrodynamics (MHD) simulation in Li et al. (2023). The test ions possess gyroradii of approximately 3.7 R$_M$ along the solar wind $\vec{E}$. For ions seeded in the magnetosheath, although they might enter the plasma mantle (Jasinski et al., 2017; Sun et al., 2020), their trajectories intersect the plasma sheet at a distance tailward of the near-Mercury neutral line, which is normally located at $X_{MSM}$ = -3 R$_M$ (Slavin et al., 2012; Sun et al., 2016; Poh et al., 2017). Ultimately, they are lost downtail due to tailward reconnection outflows.

In our study, we assume that those Na$^+$-group ions in the magnetosphere moving planetward are recycling ions (Figure 3c). The recycling rates are approximately 28% and 71% for TAA intervals of 315° to 135° and 135° to 315°, respectively, which differ significantly. However, it is essential to note that not all planetward ions are guaranteed to precipitate onto the planetary surface as some may traverse the magnetopause and enter the magnetosheath (see, Delcourt et al. 2003; Zhao et al., 2022). Therefore, these recycling rates are upper limits.

It is important to emphasize that both of the above discussed magnetospheric dynamics need to explicitly associated with the TAA in order to understand their influence on the escape of $Na^+$-group ions within TAAs.

Our study reveals that the main escape region for $Na^+$-group ions is the magnetosheath at Mercury. In contrast, $O^+$ ions primarily escape through the plasma sheet at Mars (Dong et al., 2017) and Venus (Masunaga et al., 2019), and the plasma mantle at Earth (Slapak et al., 2017). Firstly, the $Na^+$-group ions may be locally photoionized as neutrals can reach high altitude (> 1000 km) due to energetic surface release mechanisms such as micrometeoroid impact vaporization or solar wind sputtering. These neutrals can reside in the exosphere for tens of minutes, allowing a portion of elements, such as Na, Al and Si, to be photoionized (e.g., Wurz et al., 2010; Milillo et al., 2020; Jasinski et al., 2020; Moroni et al., 2023). Furthermore, ions created at low altitudes within the magnetosphere would be picked up by the magnetospheric convection flow, then a portion of these ions may be leaked out the magnetopause due to their large gyroradii and the compressed magnetopause (as simulated by Yagi et al., 2017; Glass et al., 2021).

### 4.4. Mercury's High Escape Rate of $Na^+$-group Ions Among Inner Planets

Although Mercury has a tenuous exosphere, the escape rates of $Na^+$-group ions (0.2 to $1 \times 10^{25}$ $s^{-1}$) are high compared to other inner planets in the solar system, which have much thicker atmospheres. MESSENGER's measurements were conducted between March 2011 and April 2015, coinciding the solar maximum of solar cycle 24. The $O^+$ ions are the main planetary ions in other inner planets. Their escape rates during solar maximum are 1 to $4 \times 10^{24}$ $s^{-1}$ at Venus (Masunaga et al., 2019; Persson et al., 2018), up to $10^{25}$ $s^{-1}$ at Mars (Lundin et al., 2013; Nilsson et al., 2023), $10^{25}$ to $10^{26}$ $s^{-1}$ at Earth (Slapak et al., 2017).

The high escape rate of Mercury's $Na^+$-group ions can be due to several factors. Firstly, Mercury's proximity to the Sun results in the highest solar radiation among the inner planets, promoting photoionization, and it also results in the most intense solar wind convection electric field, facilitating momentum exchange between the solar wind and pickup ions. Secondly, the $Na^+$-group ions include ions with m/q from 21 to 30 amu/e. Thirdly, the photoionization frequencies for Na ($4 \times 10^{-5}$ $s^{-1}$), Al ($6 \times 10^{-3}$ $s^{-1}$), and Si ($2 \times 10^{-4}$ $s^{-1}$) except Mg ($5 \times 10^{-7}$ $s^{-1}$), some of the elements with atomic mass from 21 to 30, at Mercury (Jasinski et al., 2020) are higher than the frequencies for O at Venus ($10^{-6}$ $s^{-1}$), Earth ($7 \times 10^{-7}$ $s^{-1}$) and Mars ($3 \times 10^{-7}$ $s^{-1}$) (Zhang et al., 1993). However, $O^+$ ions at Venus, Earth and Mars may also arise from alternative processes, including charge exchange, electron impact, and dissociative recombination of molecular ions. These processes could yield comparable rates to photoionization (Zhang et al., 1993; Lillis et al., 2017). Last but not least, within Mercury's dynamic magnetosphere, frequent magnetic reconnection deposits energy into the system causing particle and ions outflows (Sun, Dewey, et al. 2022).

Our observed peak escape rate of $10^{25}$ s$^{-1}$ for Na$^+$-group ions is much higher than the rate of $4 \times 10^{24}$ s$^{-1}$ estimated using a model considering only neutral Na (Ip, 1990) or theoretical calculations based on various intruments (Jasinski et al., 2021). The higher escape rate of Na$^+$-group ions may be attributed to: Firstly, the Na$^+$-group ions include multiple ions with m/q from 21 to 30 amu/e; Secondly, existing models did not consider sputtered ions arising from high-energy particle precipitation (see Sun, Slavin, et al., 2022). High-energy particle sputtering reaches its peak in the cusps and the nightside plasma sheet during intense magnetic reconnection intervals.

In future studies, measurements from MESSENGER and BepiColombo (Benkhoff et al., 2021), as well as simulation tools considering reconnection-generated structures in the magnetosphere, such as Hall MHD (Li et al., 2023), global hybrid (Trávníček et al., 2007; Lu et al., 2022; Guo et al., 2023) and fully kinetic simulations (Lavorenti et al., 2023), coupled to the planetary surface and exosphere, will allow for the detailed investigation of the distribution and escape of Na$^+$-group ions.

## 5. Summary

The escape of Na$^+$-group ions from Mercury was investigated by using MESSENGER's measurements. In the solar wind and magnetosheath, the ions form escape plumes with the ions concentrating in the positive solar wind $\vec{E}$ direction and the fluxes in the positive solar wind $\vec{E}$ hemisphere higher than those in the negative hemisphere.

The escape rates of the Na$^+$-group ions range approximately 0.2 to $1 \times 10^{25}$ atoms/s with the magnetosheath being the primary escape region. These escape rates are dependent on TAA with the peak rate appearing near perihelion, and are comparable during other Mercury's orbital intervals.

Despite Mercury's tenuous exosphere, its Na$^+$-group ions escape rate is similar to other inner planets, which ranges from $4 \times 10^{24}$ s$^{-1}$ at Venus to $10^{26}$ s$^{-1}$ at Earth. The high escape rate at Mercury can be attributed to multiple factors: efficient photoionization and solar wind momentum exchange due to Mercury's proximity to the Sun, frequent solar wind sputtering induced by Mercury's dynamic magnetosphere, and the inclusion of several ion species such as Na$^+$, Mg$^+$, Al$^+$, Si$^+$ etc., in Na$^+$-group ions.

**Acknowledgments**

The MESSENGER project was supported by the NASA Discovery Program under Contracts NASW-00002 to the Carnegie Institution of Washington and NAS5-97271 to The Johns Hopkins University Applied Physics Laboratory. We thank S. C. Solomon and the MESSENGER team for the successful operation of the spacecraft, and the instrument teams of FIPS from the University of Michigan, Ann Arbor, and MAG from the NASA Goddard Space Flight Center, Greenbelt, and the Johns Hopkins University Applied Physics Laboratory. We also acknowledge the team of SWMF/BATSRUS at the University of Michigan, Ann Arbor, for the simulations about Mercury's magnetosphere and test particles. Weijie Sun and James A. Slavin's contributions were supported by NASA Grants 80NSSC21K0052, Discovery Data Analysis Program Grant 80NSSC22K1061 and the National Science Foundation under Grant No. 2321595. Weijie Sun thanks Andrew Poppe (Space Sciences Laboratory, University of California, Berkeley), Yinsi Shou (University of Michigan, Ann Arbor) and Jiutong Zhao (Peking University) for their valuable discussions.

**Open Research**

The magnetic field data measured by MESSENGER are available at (Korth & Anderson, 2016). The ion data measured by FIPS onboarded on MESSENGER are available at Raines (2016). The lists of the bow shock and magnetopause used in this study are available at Sun (2023). The BATSRUS and the test particle simulations used in this study are publicly available for download as components of the Space Weather Modeling Framework at the University of Michigan (http://clasp.engin.umich.edu/swmf). The simulation output used in this study is available at Jia & Sun (2023).

# References.

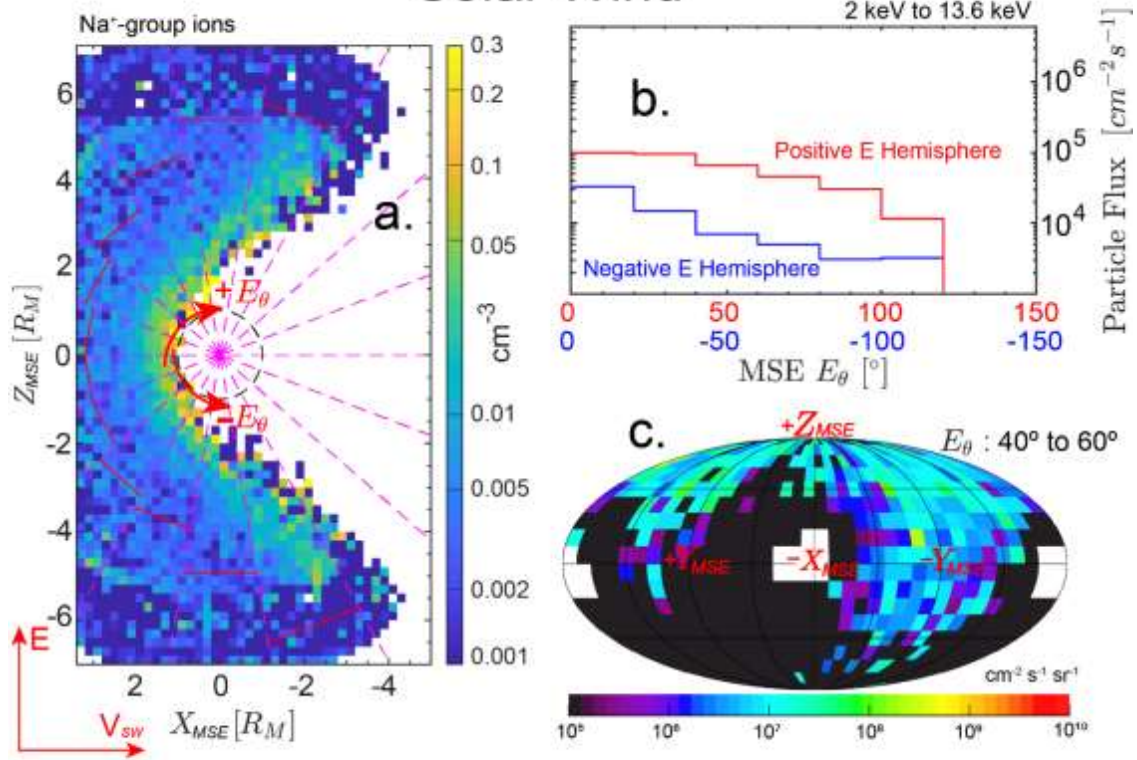

**Figure 1.** Distribution of Na$^+$-group ions in solar wind (a to c) and magnetosheath (d to f), as measured by MESSENGER from 25 March, 2011, to 30 April, 2015. (a) and (d) show the observed density ($n_{obs}$) in the $\hat{x}_{MSE} - \hat{z}_{MSE}$ plane with $|Y_{MSE}| < 1.5$ R$_M$. The $n_{obs}$ were derived by integrating the phase space density over the observed velocity range and field-of-view of FPIS (Raines et al., 2011). The solid red lines intersecting dashed lines represent the upper boundaries of FIPS measurements in the solar wind (see supplementary material for further detail). (b) and (e) show the particle fluxes along the solar wind motional electric field direction $(\vec{E})$ in the MSE $E_\theta$ intervals. $E_\theta$ is defined as atan($Z_{MSE}/X_{MSE}$). Red lines are for the $E_\theta$ intervals in +E hemisphere while blue lines are for -E hemisphere. $E_\theta$ of 180° in red aligns with -180° in blue. (c) and (f) show the 3D distribution of Na$^+$-group ions with $E_\theta$ ranging from 40° to 60°.

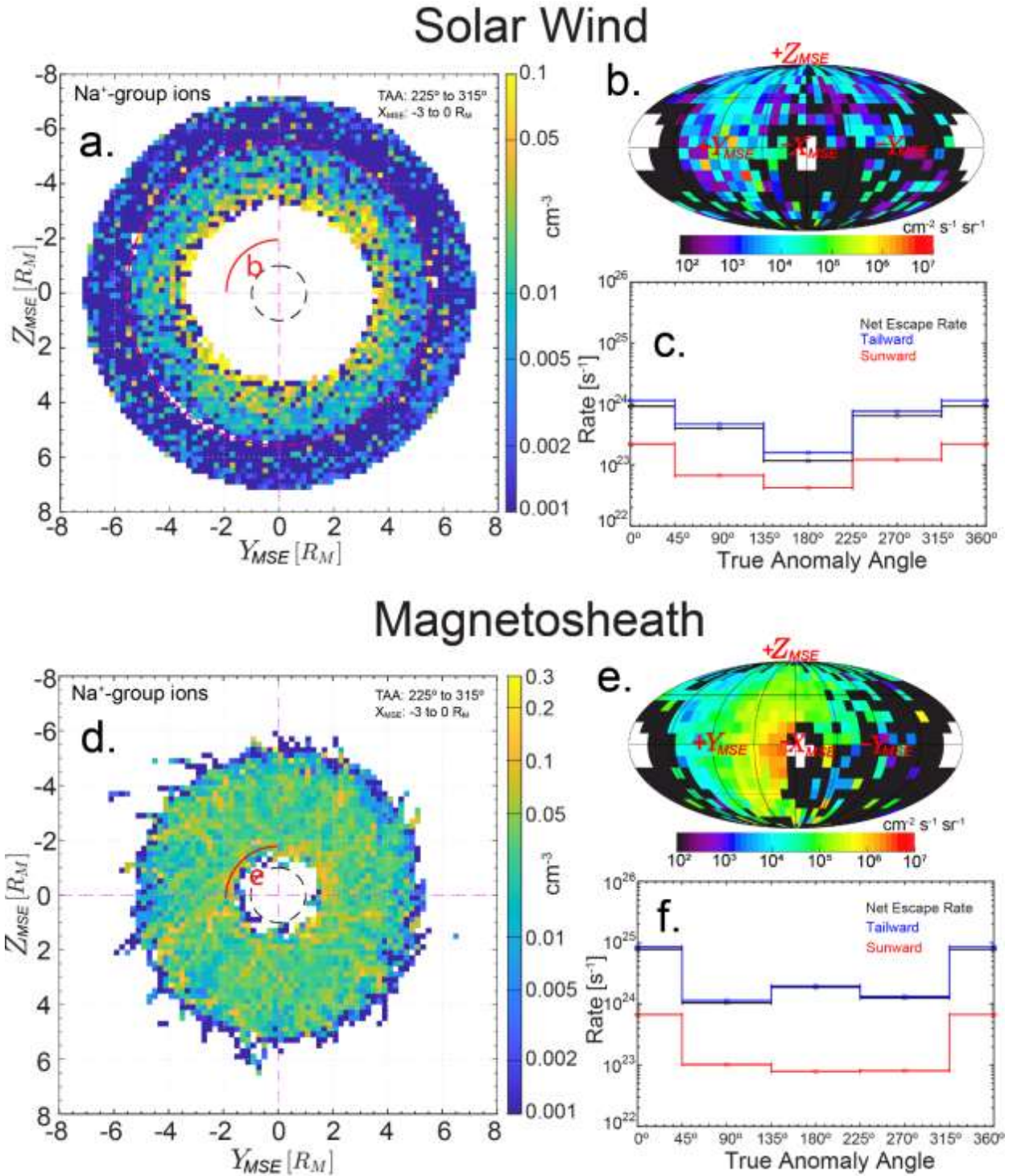

**Figure 2**. Distributions and escape rates of Na$^+$-group ions in the solar wind and magnetosheath. (a) and (d) show the cross-sections of the $n_{obs}$ in the $\hat{y}_{MSE} - \hat{z}_{MSE}$ plane with $X_{MSE}$ integrated from -3 to 0 R$_M$ and TAA from 225° to 315°. The dashed red circle in (a) represent the upper boundary of the FIPS measurements in the solar wind with a radius of 5.5 R$_M$. (b) and (e) present the integrated 3D distributions over a sector of (a) and (d), respectively. (c) and (f) show the tailward, sunward and net rates along different TAA intervals. Figures for other TAA intervals are available in the supplementary material.

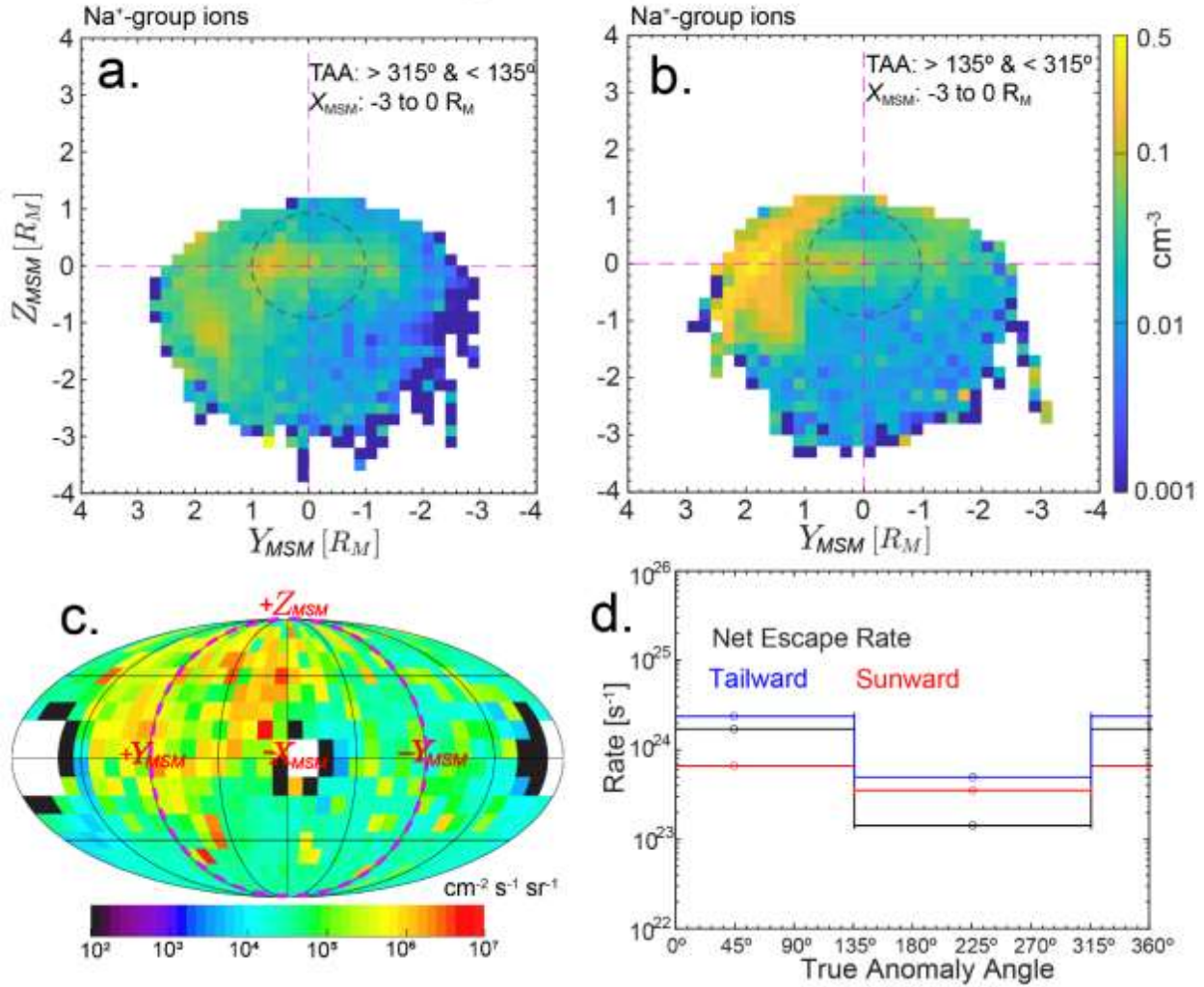

**Figure 3.** Distribution of Na$^+$-group ions in the magnetosphere with $X_{MSM}$ integrated from -3 to 0 $R_M$ and TAA from 315° to 135° (a) and 135° to 315° (b). The dashed circles indicate the planet Mercury with radius of 1 $R_M$. There are no measurements inside the planet. (c) shows the integrated 3D distribution of ions in (a). The circle in magenta encircles the tailward ions. (d) presents the tailward, sunward and net rates of the Na$^+$-group ions of the two TAA intervals.

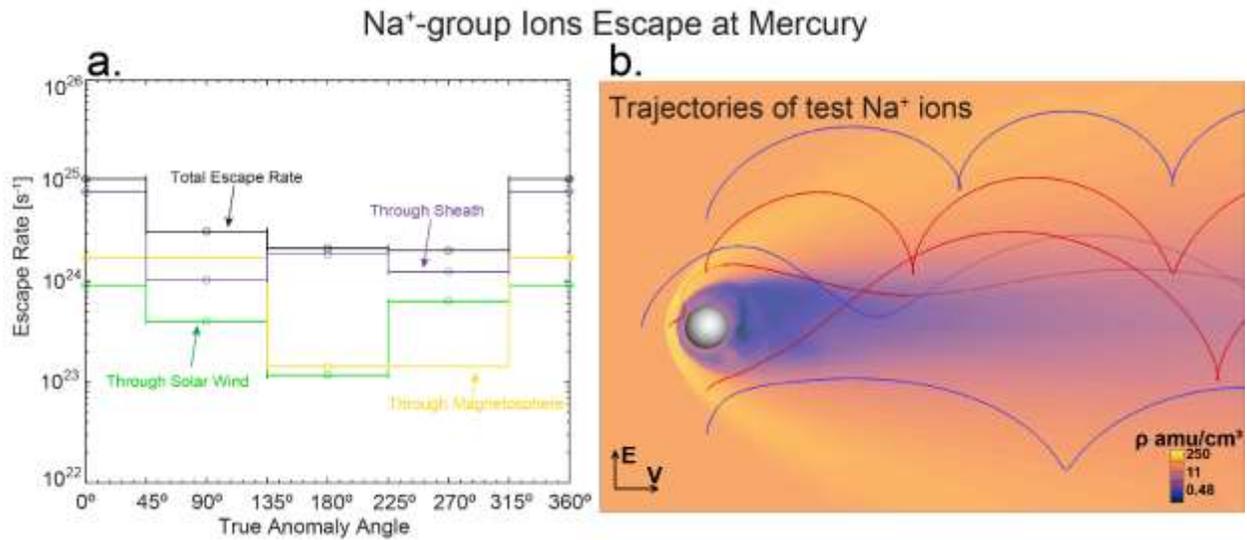

**Figure 4.** (a) The total escape rate of Na$^+$-group ions (in black), and the escape rates through solar wind (in green), magnetosheath (in purple) and magnetosphere (in yellow) along TAA intervals. (b) The trajectories of test Na$^+$ ions seeded in the magnetosheath (in red) and solar wind (in blue). These trajectories are traced the global electromagnetic fields extracted from the BATSRUS global Hall-Magnetohydrodynamics (MHD) simulation (Tóth et al., 2008). The Hall-MHD simulation run used here is from Li et al. (2023).